\documentclass[runningheads]{llncs}
\usepackage{geometry}
\usepackage{fontawesome5}
\usepackage{tikz}
\usepackage{rotating}
\usepackage{pgf-pie} 
\usepackage{pgfplots} 
\usetikzlibrary{chains,positioning,arrows.meta,quotes,shapes,fit}
\usepackage{listings}
\usepackage{subcaption}
\usepackage{graphicx}

\newsavebox{\measurebox}
\usepackage{hyperref}

\usepackage{booktabs,wrapfig}
\usepackage{tabularx}
\newcolumntype{b}{X}
\newcolumntype{a}{>{\hsize=.08\hsize}X}
\newcolumntype{t}{>{\hsize=.2\hsize}X}
\newcolumntype{k}{>{\hsize=.4\hsize}X}
\newcolumntype{s}{>{\hsize=.6\hsize}X}
\usepackage{footnote}
\makesavenoteenv{tabular}
\makesavenoteenv{table}

\begin{document}
\title{User Interaction Data in Apps: \\ Comparing Policy Claims to Implementations\thanks{\textit{Published at the 18th IFIP Summer School on Privacy and Identity Management 2023 (IFIPSC 2023).}}}
\titlerunning{Comparing Policy Claims to Implementations}
\author{Feiyang Tang \orcidID{0000-0002-8720-6743} \and
Bjarte M. \O stvold \orcidID{0000-0001-6922-4027}}
\authorrunning{F. Tang and B.M. \O stvold}
\institute{Norwegian Computing Center\\N-0314 Oslo, Norway \\
\email{\{feiyang,bjarte\}@nr.no}}
\maketitle              % typeset the header of the contribution
\begin{abstract}
%As mobile app usage continues to rise, so does the generation of extensive user interaction data, which includes actions such as swiping, zooming, or time spent on a screen. Despite the significant insights that can be learned from such data, it is often collected by apps and their services without sufficient disclosure in their privacy policies. A common issue is that many apps classify this data as non-personal, a stance that is controversial given its potential to reveal personal details when aggregated.

%In response to this issue, we propose an automated approach to check a privacy policy in the following respect: We compare the policy's claims about the app's collection of user interaction data to the actually implemented collection through static analysis of the app. This process allows us to identify inconsistencies in the claims and also to study general collection practices for user interaction data across apps.
%Via an improved comparison between data collection claims and actual implementations, our approach aims to enhance transparency, foster trust between app developers and users, and contribute to a more informed discussion on the classification of user interaction data.

As mobile app usage continues to rise, so does the generation of extensive user interaction data, which includes actions such as swiping, zooming, or the time spent on a screen. Apps often collect a large amount of this data and claim to anonymize it, yet concerns arise regarding the adequacy of these measures. In many cases, the so-called anonymized data still has the potential to profile and, in some instances, re-identify individual users. This situation is compounded by a lack of transparency, leading to potential breaches of user trust. 

Our work investigates the gap between privacy policies and actual app behavior, focusing on the collection and handling of user interaction data. We analyzed the top 100 apps across diverse categories using static analysis methods to evaluate the alignment between policy claims and implemented data collection techniques. 
Our findings highlight the lack of transparency in data collection and the associated risk of re-identification, raising concerns about user privacy and trust.
This study emphasizes the importance of clear communication and enhanced transparency in privacy practices for mobile app development.

\keywords{Mobile Apps \and Transparency \and Trust \and Interaction Data \and Privacy Policy}
\end{abstract}

\section{Introduction}
Mobile apps have become deeply integrated into daily life, often collecting user interaction data like taps and swipes. While this data is typically anonymized to protect privacy, the effectiveness of this anonymization is increasingly under scrutiny. Anonymized data, when aggregated, can still enable user profiling and potentially lead to identification. This challenges the common practice of labeling such data as ``non-personal'' to meet less stringent privacy regulations. These practices, under the pretext of anonymization, pose significant privacy risks and can diminish user trust. 

To address these issues we propose an automated method to compare privacy policy statements with actual data collection practices in app code. We aim to highlight the discrepancies between policy and practice, thereby enhancing transparency and rebuilding trust. Our focus extends beyond mere regulatory compliance to advocating for stronger data protection and a culture of transparency in the digital domain.

This paper aims to answer the following research questions:
\begin{enumerate}
\item What claims do app privacy policies make concerning the collection of user interaction data?
\item What insights can be derived from analyzing app implementations in light of policy claims?
\item How can we automate the examination of the transparency of collection claims in privacy policies based on evidence obtained by static analysis?
\end{enumerate}
Our contributions extend and automate our previously work~\cite{tang2023transparency}:
\begin{enumerate}
\item Extending the manual analysis approach~\cite{tang2023transparency}, we introduce an automated claim extractor and classifier for processing privacy policies. This approach uses natural language processing techniques, enhanced by targeted keyword searches, to automatically extract and categorize claims about user interaction data collection.
\item We develop a static analyzer and an evidence classifier. These components automatically extract and categorize details of user interaction data collection directly from app implementations, streamlining the process.
\item By automating the comparison of labeled collection claims (extracted from privacy policies) with the labeled collection evidence (derived from application code), our approach provides a more efficient and objective assessment of the transparency of data collection practices.
\item Building upon the automated components, we conduct a study of 100 popular mobile apps. This study aims to analyze and identify patterns in user interaction data collection, enhancing the understanding of this practice and its implications for privacy and transparency.
\end{enumerate}

\begin{figure}[htp]
\centering
\tikzstyle{block} = [rectangle, fill=gray!20,draw,node distance=2cm,
text width=6em, text centered, rounded corners, minimum height=4em]
\tikzstyle{cloudblu} = [draw, ellipse,fill=blue!20, node distance=2cm,
minimum height=4em, text width=6em, text centered]
\tikzstyle{cloud} = [draw, ellipse,fill=red!20, node distance=2cm,
minimum height=4em, text width=6em, text centered]
\resizebox{\textwidth}{!}{%
\begin{tikzpicture}[node distance=15mm and 25mm,box/.style = {draw, minimum height=15mm, text width=25mm, align=center},sy+/.style = {yshift= 2mm}, sy-/.style = {yshift=-2mm},every edge quotes/.style = {align=center}]
\node (n1) [block]             {Privacy policy};                  
\node (n2) [cloud,right=of n1] {Claim extraction};
\node (n3) [cloud,right=of n2] {Claim classification};  
\node (n5) [block,below=of n1] {APK file};
\node (n6) [cloudblu,below=of n2] {Static analysis};
\node (n7) [cloud,below=of n3] {Evidence \\categorization};
\path (n3) -- (n7) coordinate[midway] (aux);
\node (n9) [cloudblu,right=of aux,node distance=5cm] {Fact-check};
%%% RECTANGLES %%%
\node[draw, thick, dotted, rounded corners, inner xsep=1em, inner ysep=1em, fit=(n2)(n3)] (box) {};
\node[fill=white] at (box.south) {NLP};
%%% RECTANGLES %%%
\node[draw, thick, dotted, rounded corners, inner xsep=1em, inner ysep=1em, fit=(n1)(n5)] (box) {};
\node[fill=white] at (box.center) {Mobile app};
\draw[thick,->]  (n1.east) to [above,"policy\\text"] (n2.west);
\draw[thick,->]  (n2.east) to [above,"policy\\segment"] (n3.west);
\draw[thick,->]  (n3.east) to [above right,"labeled\\collection claim"] (n9.north);
\draw[thick,->]  (n5.east) to [above,"bytecode\\+ API list"] (n6.west);
\draw[thick,->]  (n6.east) to [above,"invocation\\+UI file"] (n7.west);
\draw[thick,->]  (n7.east) to [below right,"labeled\\collection evidence"] (n9.south);
\end{tikzpicture}%
}
\caption{Overview of the approach}
\label{fig:flowchart}
\end{figure}
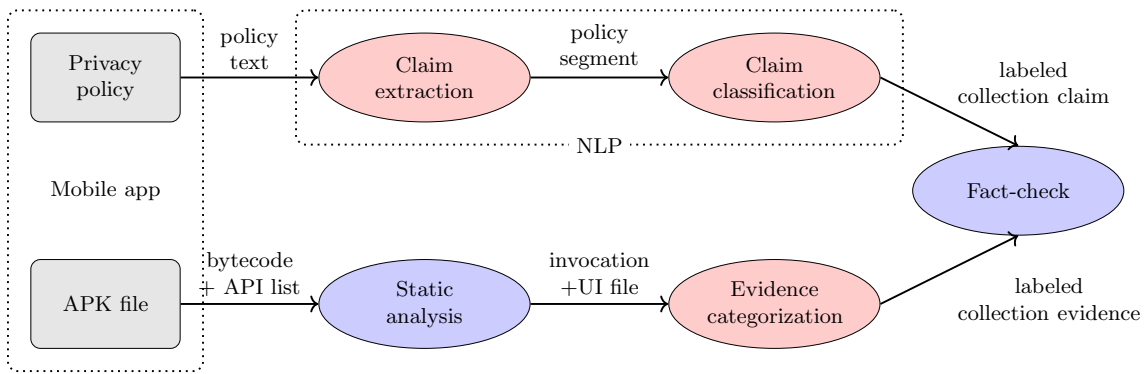

Our two-fold approach, encompassing privacy policy analysis and application code analysis, is depicted in Fig.~\ref{fig:flowchart}.

\section{Motivation}
The capability of anonymized user interaction data to be de-anonymized and thus potentially classified as personal data under regulations such as the GDPR is an emerging concern in user privacy. Studies, including those by Leiva et al.~\cite{leiva2021my}, highlight that even data devoid of explicit personal identifiers can be subjected to user profiling and identification of users through the aggregation of interaction data with other contextual information. This complex interplay blurs the distinction between non-personal and personal data, thus challenging the notion that anonymized data is inherently non-sensitive or non-identifiable.

Furthermore, the work of Cre{\c{t}}u et al.~\cite{crectu2022interaction} suggests that stable behavioral patterns within anonymized mobile app data can be leveraged to achieve high re-identification rates. Such findings directly challenge the GDPR's current classification of this data as non-personal. This misinterpretation of what constitutes personal data, especially in the context of user interaction, can significantly undermine privacy risks. It also raises questions about the effectiveness of anonymization techniques and the need for informed consent and transparent data governance, even when data is seemingly anonymized~\cite{grunewald2021tilt}.

The ambiguity in categorizing data as personal or non-personal is further complicated in mixed datasets, where differentiating between the two becomes increasingly challenging~\cite{bp1}. This is particularly relevant in the context of mobile apps, where user interaction data is often collected alongside other types of data. The risk of re-identification in what is classified as non-personal data underscores the importance of rethinking how such data is treated within legal frameworks. As argued in existing literature, there is a growing necessity to treat non-personal data with the same level of protection as personal data~\cite{bp2}.

Transparency in mobile app data collection is critical for user trust and app adoption, as it ensures user autonomy and accountability~\cite{tang2023transparency}, especially with the rise of analytics services that may infringe upon GDPR guidelines~\cite{SeveralE79:online}; thus, fostering transparency and user control is imperative for sustaining user satisfaction and promoting app engagement.

\section{Analyzing Collection Claims from Privacy Policies}

In an attempt to assess the transparency of data collection practices stated in privacy policies, we developed a two-tiered approach. This strategy is specially designed to extract and classify claims related to user interaction data collection, a facet less explored in privacy policy analysis.

This approach aims to answer three primary questions:
\begin{itemize}
\item \textit{Does the privacy policy mention user interaction data collection?}
\item \textit{If so, what types of user interaction data are claimed to be collected?}
\item \textit{What techniques are claimed to be used for this data collection?}
\end{itemize}

\subsection{Claim Extraction}
The first phase of our approach identifies whether user interaction data collection is mentioned within a privacy policy. Instead of conventional keyword-based approaches, this extractor utilizes semantic context to accommodate the diverse ways such claims can be articulated.

The APP-350 Corpus was utilized in this stage \cite{zimmeck2019maps}. This corpus comprises 350 Android app privacy policies annotated for privacy practices. However, the existing annotations primarily focused on personal data collection, which didn't coincide with our emphasis on user interaction data collection. Therefore, we conducted our own manual annotations.
\newline

\noindent\textbf{Key Findings:}
Our review of the 350 app privacy policies yielded several key findings that offered insights into the disclosure practices regarding user interaction data collection.

%\begin{wraptable}{r}{0.4\textwidth}
\begin{table}[t]
\centering
\caption{Most frequent bigrams}
\label{tab:bigrams}
\begin{tabular}{lr}
\toprule
\textbf{Bigram} & \textbf{Freq.} \\ \midrule
Your Information & 3,295 \\
Our Service & 2,941 \\
Your Data & 2,892 \\
Third Party & 1,788 \\
Help You & 1,214 \\
Improve Service & 947 \\
Automatic Collection & 422 \\
Tracking Technology & 402 \\
Interact With & 346 \\
Collect Information & 281\\
\bottomrule
\end{tabular}
\end{table}
%\end{wraptable}

By doing manual annotation we found that out of the 350 analyzed apps, 294 mentioned the collection of user interaction data at varying detail levels. However, of these 294, only 57\% (169/294) of the policies provided more specifics than a mere mention of ``data'' or ``information.''
Upon segmenting the privacy policies into sentences, we annotated 3,661 sentences as relevant to user interaction data collection from a total of 42,797 sentences.
Table \ref{tab:bigrams} presents the most frequently occurring bigrams within these annotated sentences.

Transparency about user interaction data collection varied significantly across apps. Although 294 policies referenced such data collection, the details were often obscured by general phrases like ``\emph{we collect data to improve our service}.''
Our bigram analysis highlighted the common use of third-party services in the data collection process. These services, often referred to as ``\emph{tracking technology}'', are employed to automatically collect data purportedly to enhance services. ``\emph{Google Analytics}'', a prominent third-party analytics service, was frequently observed in our bigram analysis, underscoring its vital role in user interaction data collection.

%Sec~\ref{Sec:claimimple} will provide further details on the dataset analysis and the training and testing process of the claim extractor.

\subsection{Claim Classification}

The second phase of our approach classifies the claims. This model is innovative in its ability to categorize claims according to user interaction data types and collection techniques. Unlike traditional binary classifiers, it acknowledges that a single sentence may convey multiple types of data and collection techniques.
\newline

\noindent\textbf{Key Findings:}
In our examination of the 169 privacy policies that offered more explicit information about user interaction data collection, we found that only 56 policies clearly stated the collection techniques, such as \emph{``the times you click a page''} or \emph{``the time you spend watching content''}.

To standardize vocabularies and taxonomy for classification purposes, we utilized data types and collection techniques from our previous work, known as collection vocabularies~\cite{tang2023transparency}.
These vocabularies included six types of interaction data and an additional category named device data, which we observed is commonly collected alongside interaction data. The frequencies with which of different data collection types are mentioned in the policies are shown in Table~\ref{tab:datacollection}.

\begin{table}[t]
%\begin{wraptable}{r}{0.4\textwidth}
\centering
\caption{Data collection frequencies (169 apps)}
\label{tab:datacollection}
\begin{tabular}{lr}
\toprule
\textbf{Data Type} & \textbf{Freq.} \\ \midrule
App Presentation & 98\% \\
Categorical & 60\% \\
User Input & 45\% \\
Binary & 17\% \\
Gesture/Composed Gesture & 2\% \\
\textit{Device Data} \footnote{While not part of interaction data, it is a crucial component often collected alongside interaction data. This includes information such as the International Mobile Equipment Identity (IMEI) number, device model, operating system, and other device-specific identifiers.} & 92\% \\
\bottomrule
\end{tabular}
\end{table}
%\end{wraptable}

The descriptions given by the apps about their collection techniques were often vague. Of the 56 apps that vaguely mentioned the techniques used, all referred to frequency, representing 100\% of this subgroup. A substantial but smaller portion, 48\% (27 out of 56), mentioned duration, using phrases like \emph{``time spent watching''} or \emph{``length of service use''}. However, only a mere 1.8\% (1 out of 56) of these apps mentioned motion.

Transparency was lacking in the descriptions of user interaction data collection types and techniques. Of the 294 apps that acknowledged data collection, a majority, 84\% (248 out of 294), categorized the collected data as \emph{``non-personal data''}, without providing further details. 
Such categorization seemed to be used to justify sensitive actions like \emph{``aggregation''}, a method mentioned by 43\% (126 out of 294) of these apps, and \emph{``transfer to third-party services''}, an action mentioned by 68\% (199 out of 294). 
Furthermore, almost half, 48\% (141 out of 294), acknowledged using \emph{``automatic collection''} methods.

Further details on dataset analysis, along with the training and testing procedures for both the claim extractor and claim classifier, can be found in the Appendix under Section~\ref{Sec:claimimple}.

\section{Analyzing Collection Evidence from Application Code}

Once the collection claims from the privacy policy have been extracted, we seek to validate these claims by investigating the application code for tangible signs of data collection. Our attention is primarily devoted to identifying and categorizing the embedded data collection techniques within the mobile app.
The approach we adopt for static analysis explicitly targets user interface (UI) elements and the invocations to analytics libraries from these UI elements. Following identification, these elements are classified based on a predefined collection vocabulary that we have introduced in~\cite{tang2023transparency}. This vocabulary was generated through a meticulous examination of all Android UI widgets and it captures a broad range of user interaction data types and collection techniques.
The detailed terms for types of user interaction data and collection techniques employed in our study are listed in the Appendix.

The collection vocabulary not only allows for a structured classification of data collection instances but also facilitates the mapping between the collection evidence found in the code and the claims made in the privacy policy. The usage of this comprehensive vocabulary ensures that we can conduct a granular comparison later in the fact-checking process.

Through our analysis, we aim to answer the following questions:
\begin{itemize}
\item \emph{Which analytics libraries are being utilized by the mobile app?}
\item \emph{What types of user interaction data are being collected?}
\item \emph{Which techniques are employed for data collection?}
\end{itemize}

\subsection{Analytics Library Identification}
In the first stage of our code analysis, we focus on identifying the analytics libraries that are used by the mobile apps. It is common for apps to utilize such libraries to gather and analyze user interaction data, providing developers with valuable insights into user behavior.

To achieve this, we target a set of popular analytics libraries as our initial point of focus. These libraries are often integral to tracking user interactions and facilitating data collection. Hence, recognizing these libraries' invocations serves as an efficient guide to pinpoint locations where user interaction data collection is likely to take place.

Our analysis primarily focuses on the classes that engage with UI elements, carefully examining the imported analytics libraries along with their respective method invocations. We constrain our investigation to a selected set of methods belonging to popular analytics libraries that are frequently utilized for data collection. In this context, we adopted the list of the top 20 analytics services for Android apps listed on AppBrain\footnote{\url{https://www.appbrain.com/stats/libraries/tag/analytics/android-analytics-libraries}}. Prior understanding of these frequently used analytics libraries and their APIs forms a crucial foundation for this stage of our analysis.

\subsection{Categorizing Data Types and Collection Techniques}
Following the identification of analytics libraries, our objective is to establish links between the UI elements and the corresponding bytecode that manages user interactions. UI actions such as button presses trigger specific methods within the bytecode. Thus, we delve into both XML files, which define the UI elements, and the bytecode, which dictates the actions corresponding to these elements. The examination of these components often provides insights into the type of user interaction data being collected.

For instance, consider a simple scenario where a Firebase Analytics library is employed in an Android app. A button click in the UI represented as \texttt{<Button android:onClick="buttonClick"/>} in the XML file, would trigger a corresponding \texttt{buttonClick(View view)} method in the Java code. 
The interaction with the analytics library within this method could look something like this:

\begin{lstlisting}[basicstyle=\ttfamily\footnotesize]
public void buttonClick(View view) {
    FirebaseAnalytics mFAnalytics = FirebaseAnalytics.getInstance(this);
    Bundle params = new Bundle();
    params.putString("Button_name", "button1");
    params.putString("Action", "click");
    mFAnalytics.logEvent("ButtonClick", params);
}
\end{lstlisting}

Here, an invocation to the Firebase Analytics library occurs whenever the button is clicked, recording the button's name and the associated action. This example highlights that click data is collected each time the button is clicked.

Though this method generally proves effective in discerning the types of user interaction data being collected, it is important to note that some complexities in the bytecode may obscure certain data collection events. Additionally, data collected outside of standard UI interactions, such as device-generated data or data from non-UI sources, may not be captured by this approach.
Building upon the successful linking of UI elements to their corresponding analytics library invocations, we categorize the extracted data based on predefined interaction data types and collection techniques. Our initial focus is on the types of user interaction data, where we aim to classify the data according to their corresponding UI elements. 
Table~\ref{tab:UIElements} presents a classification of interaction data types associated with common Android UI elements.

\begin{table}[t]
\centering
\caption{Types of user interaction data and corresponding UI elements}
\label{tab:UIElements}
\begin{tabularx}{\textwidth}{sb}
\toprule
\textbf{Interaction Data Types} & \textbf{Android UI Elements} \\
\midrule
App Presentation & View (TextView, VideoView, WebView, etc.) \\ \midrule
Binary & Button (ImageButton, CheckBox, etc.) \\ \midrule
Categorical & AbsSpinner (Spinner), CompoundButton (RadioButton, Switch), RatingBar \\ \midrule
User Input & TextView (EditText, AutoCompleteTextView, SearchView) \\ \midrule
Gesture & GestureDetector, ViewPager, SwipeRefreshLayout \\ \midrule
Composite Gestures & GestureDetector (ScaleGestureDetector) \\
\bottomrule
\end{tabularx}
\vspace{-5mm}
\end{table}
In the table, the main Android UI elements represent the core classes or interfaces in the Android UI hierarchy. For instance, View is a fundamental class for UI widgets in Android, and the various UI elements like TextView, VideoView, and WebView are its subclasses, hence included as its subcategories. 

In this process, we perform an inspection of each UI element across the XML files, which define the UI, and the code files that handle these UI elements. Accordingly, the type of user interaction data is ascertained based on the functionality attributed to the UI elements.

\subsubsection{Identification of Collection Techniques}

Our approach to identifying the collection techniques for user interaction data consists of two components: rules-based identification using predefined criteria, and criteria obtained from a detailed analysis of popular analytics libraries' documentation.

In rules-based identification, we create a set of heuristics centered on invocations of Android or Java methods, which are associated with different collection techniques.
For instance, the ``frequency'' technique can be inferred from the event logging invocation. Techniques like ``duration'' collection can be suggested by invocations of methods from the Java \texttt{Timer} class or \texttt{android.os.SystemClock.elapsedRealtime()}. Similarly, ``motion details'' collection stem from methods in the \texttt{MotionEvent} class, such as \texttt{getPressure()}, \texttt{getX()}, and \texttt{getY()}.

The second component of our approach involves using criteria obtained from the documentation of widely-used analytics libraries, such as Firebase Analytics and Mixpanel. Once the specific API methods used for different collection techniques in these libraries are identified, they are added to our categorization list. For instance, Firebase Analytics' \texttt{logEvent()} method, with parameters like \texttt{select\_content} and \texttt{view\_item}, can log the frequency of user interactions. On the other hand, Mixpanel uses the \texttt{track()} method with event names to record frequency. For recording duration, Firebase Analytics uses the \texttt{user\_engagement} event, capturing user engagement duration, while Mixpanel provides the \texttt{time\_event()} method to time events' duration.

While this approach provides a systematic and informed means to identify collection techniques, it also has limitations. For example, if an app uses a custom package without Java or Android method invocations, or if it uses a third-party service not included in our list, our categorization method may not accurately identify the collection technique used. Further details on the performance metrics are provided in the Appendix under Section~\ref{Sec:evidencemetrics}.

\section{Fact-Checking Privacy Policy Claims}\label{Sec:factcheck}

Upon completing the static analysis and organizing the privacy policy collection claims, we have the necessary foundation to perform a fact-checking analysis on these claims. The goal of this process is to detect any inconsistencies between the data collection practices described in the policy and the actual practices observed in the application code. The process unfolds in two stages:

\subsection{Mapping Interaction Data Types and Collection Techniques}
In the first stage, we create a mapping between the types of data outlined in the privacy policy and the equivalent interaction data types pinpointed during our static analysis. A similar mapping is constructed for each collection technique stated in the policy and the corresponding technique identified within the application code.

For instance, suppose a privacy policy declares, ``\emph{We collect the content you provide}'', implying the collection of user-input data. During our static analysis, we identify the invocation of \texttt{EditText} elements in the application code, which signifies user input in Android. We then form a mapping between the phrase ``\emph{We collect the content you provide}'' from the privacy policy and the \texttt{EditText} elements found in the code.

In another case, if the policy statement indicates, ``\emph{We track how long you spend on our services}'', this suggests the usage of a duration-based collection technique. Suppose we identify the invocation of \texttt{android.os.SystemClock.elapsedRealtime()} in the code, which measures elapsed time, a mapping is established between the policy phrase ``\emph{We track how long you spend on our services}'', and the this invocation in the code.

These mappings provide a basis for comparing the privacy policy's claims to the actual evidence in the code, allowing us to assess the consistency between policy declarations and the application code's actual practices.

\subsection{Interaction Consistency Analysis}

Having established the mappings, we can compare the data types and collection techniques from the privacy policy to those discovered in the code. This allows us to calculate the \textbf{Interaction Consistency Rate}, which measures the extent of consistency between the collection evidence identified in the static analysis (categorized by data type and collection technique) and the corresponding claims in the privacy policy. This rate represents the proportion of collection evidence found in the code that is accurately claimed in the policy.

An inconsistency may arise if, for example, our static analysis uncovers \texttt{EditText} invocations, but there is no mention of ``user input data'' in the app's privacy policy. Note that our analysis focuses on correlating claims made in the privacy policies with evidence gleaned from our static analysis. This means that if data collection is linked with a UI element that falls outside the scope of our static analysis, such collection will not be included in our investigation.

\subsection{Context Consistency Analysis}

The second stage of our analysis involves a context-based examination to comprehend when user interaction data is collected. Our motivation for conducting a context-based analysis is based on our preliminary observation from the APP-350 dataset, where 74\% of the policy sentences related to user interaction data collection also described the context, for example, ``We collect information on how you interact with our service \emph{when you are making a purchase}.''

To accomplish this, we review the app's code and identify unique contexts under which data collection takes place. The contexts we consider here are confined to those directly linked with identifiable criteria in the bytecode, thereby limiting our scope to certain discernible contexts.

Our approach for constructing a context catalog began with a careful selection of Android apps. We chose a representative sample of 25 apps from five distinct categories within the Google Play Store in Germany. \footnote{The German Google Play Store was selected for its adherence to the GDPR, ensuring that the apps included in the study would have well-constructed privacy policies. \url{https://play.google.com/store/apps?hl=en_US&gl=DE}} 
The categories selected were: ``Social Networking'', ``Health \& Fitness'', ``Entertainment'', ``Productivity'', and ``Finance''. These categories were chosen for their popularity and the likelihood that they would handle a decent amounts of user interaction data.
Each of these apps underwent a detailed static analysis. We scanned their bytecode for instances of user interaction data collection, focusing on the specific contexts in which this collection occurred. 

Through this process, we identified and organized recurring contexts across the different apps. These common contexts, indicative of typical scenarios associated with user interaction data collection are developed into a generalized catalog. While not comprehensive, this catalog, as presented in Table~\ref{table:contexts}, provides an informative overview of the most common user actions and apps states where interaction data collection is likely to occur.

\begin{table}[t]
\centering
%\scriptsize
\caption{Catalog of contexts for user interaction data collection}
\label{table:contexts}
\begin{tabularx}{\textwidth}{kb}
\toprule
\textbf{Context} & \textbf{Identifiable Criteria in Code} \\
\midrule
Viewing Content & Invocation of certain View UI elements (e.g., TextView/ImageView). \\\midrule
Making Purchase & Calls to Android Google Play payment service APIs. \\\midrule
Location-Based Services & Invocation of Android Location APIs. \\\midrule
Interacting with Media & Calls to media-related APIs (e.g., Media Player, Media Recorder). \\\midrule
Search & Invocation of SearchView UI elements. \\\midrule
Notifications & Interactions with NotificationManager API. \\\midrule
Accessing User Profile & Invocation of User Profile related APIs (e.g., AccountManager). \\\midrule
Sensor-based Features & Use of Android Sensor APIs. \\\midrule
Communication Features & Use of communication-related APIs (e.g., TelephonyManager). \\\midrule
Gameplay Interactions & Calls to APIs related to gameplay, typically seen in game apps. \\\midrule
Customization Features & Invocation of APIs related to customization (e.g., changing theme). \\
\bottomrule
\end{tabularx}
\vspace{-5mm}
\end{table}

Based on this catalog, we calculate the \textbf{Context Consistency Rate}, which measures the degree of consistency between the data collection contexts identified in the static analysis and those outlined in the privacy policy. This rate indicates the proportion of collection contexts found in the code that are accurately represented in the policy. 

We recognize that our catalog cannot encapsulate all possible contexts due to the complexity and diversity of user interactions and app functionalities. Furthermore, our policy claim checks rely on language model-assisted vocabulary matching, which might not guarantee absolute precision. These factors should be considered when interpreting the Context Consistency Rate.

\section{Experiment}
In this section, we present the results of a large-scale analysis conducted on a set of 100 Android apps. Through this comprehensive examination, we aim to gain insights into the landscape of user interaction data collection practices as reflected in their privacy policies and underlying code. This analysis forms the basis of our discussion on the consistencies and discrepancies between policy claims and actual code execution.

\subsection{Setup}

Our experimental analysis is based on a set of Android apps obtained from the Google Play Store in Germany. To ensure a comprehensive and varied dataset, we selected the top 100 apps from 10 distinct popular categories. These categories included varied domains such as ``Lifestyle'', ``Education'', ``Travel'', and 'Entertainment' among others, chosen for their relevance to a broad spectrum of users and potential data collection diversity.

We employed two key criteria for selecting these apps: (1) The categories and apps should be disjoint to avoid overlap and redundancy in our dataset. This approach was crucial to ensure that each app provided unique insights into user interaction data collection practices. (2) Every app must have a corresponding English privacy policy webpage linked in its ``Data Safety'' section. This criterion was essential to facilitate the analysis of privacy policies against actual app behaviors, and to ensure that all apps adhered to the GDPR. The chosen apps represented a mix of global popularity and regional relevance.%, reflecting a balance between widely-used applications and those catering to specific user needs within Germany.

In this experiment, we assess the data's consistency from privacy policies against static analysis results using two primary metrics: the Interaction Consistency Rate and the Context Consistency Rate, detailed in Section~\ref{Sec:factcheck}. 
These metrics measure the alignment of data types and collection contexts between policy claims and code evidence. 
We also introduce the \textbf{Interaction Consistency Coverage Rate} and the \textbf{Context Consistency Coverage Rate} to determine the completeness of our analysis, identifying any potential gaps in our static analysis method. These coverage rates help pinpoint areas not covered by our analysis, ensuring a thorough evaluation of privacy policy claims.

\subsection{Overview of User Interaction Data Collection Practices}

%Firstly, we provide an overview of the user interaction data collection practices as stated in the privacy policies and as evidenced in the application code of the 100 analyzed apps.
%
In our overview of 100 apps, illustrated in Fig.~\ref{fig:claimresult}, indicate that 14\% of the apps do not mention any form of user interaction data collection in their privacy policies. Approximately a third of the apps (29) acknowledge data collection but do not specify the type of data collected or the method of collection. These policies often contain general statements such as, ``we use statistical tools to collect non-personal data such as usage details.'' It is important to note that the more detailed policies tend to describe the types of data collected more than the methods of collection.

%Upon analyzing the alignment between the privacy policies and the app code, we observed that the average Interaction Consistency Rate was 58\%. This rate measures the consistency between the types of interaction data collection stated in the policies and those found in the apps. The Context Consistency Rate, on the other hand, averaged at 32\%, assessing the match between the context of data collection as described in the policies and our findings in the app code. These metrics indicate that only about half of the interaction data collection practices and about a third of the collection contexts are clearly communicated to users. This finding underscores a clear gap in transparency regarding user interaction data collection practices in mobile applications. 
%The Interaction Consistency Coverage Rate stood at 86\%, while the Context Consistency Coverage Rate reached 71\%. These figures suggest that our static analysis successfully captures a large portion of interaction data collections and around two-thirds of the contexts in which these data collections occur.
Our analysis showed an Interaction Consistency Rate of 58\%, indicating how often app behaviors matched their policy claims regarding collected data types. The Context Consistency Rate was 32\%, reflecting how well the context of data collection in the app code aligns with policy claims. These rates reveal a clear gap in transparency, with our static analysis capturing most data collection instances and contexts, evidenced by an Interaction Consistency Coverage Rate of 86\% and a Context Consistency Coverage Rate of 71\%.

\begin{figure}[htp]
\centering
\begin{tikzpicture}[scale=0.6]
\tikzstyle{every node}=[font=\footnotesize\bfseries]
    \pie[
        color = {
            yellow!95!black, 
            %green!60!black, 
            blue!20, 
            red!50,
            gray!70,
            teal!30},
            explode = 0.25,
        text = legend
        ]
        {9/Both type and technique,
        41/Data type only,
        7/Collection technique only,
        29/Mention the collection only,
        14/None}
\end{tikzpicture}
\caption{Policy claims completeness regard to interaction data collection}
\label{fig:claimresult}
\vskip\floatsep% normal separation between figures
\begin{minipage}{0.49\textwidth}
\centering
\begin{tikzpicture}[scale=0.50]
\tikzstyle{every node}=[font=\LARGE\bfseries]
  \begin{axis}[
    xbar,
    y axis line style = { opacity = 0 },
    axis x line       = none,
    tickwidth         = 0pt,
    enlarge y limits  = 0.2,
    %enlarge x limits  = 0.02,
    symbolic y coords = {Presentation, Binary, Categorical, Input, Gesture},
    nodes near coords,
  ]
  \addplot coordinates {(100,Presentation)(84,Binary)(39,Categorical)(65,Input)(17,Gesture)};
  \legend{\% of apps}
  \end{axis}
\end{tikzpicture}
\caption{Data type distribution}
\label{fig:evidence1}
\end{minipage}\hfill
\begin{minipage}{0.49\textwidth}
\centering
\begin{tikzpicture}[scale=0.50]
\tikzstyle{every node}=[font=\LARGE\bfseries]
  \begin{axis}[
    xbar,
    y axis line style = { opacity = 0 },
    axis x line       = none,
    tickwidth         = 0pt,
    enlarge y limits  = 0.33,
    %enlarge x limits  = 0.02,
    symbolic y coords = {Frequency, Duration, Motion},
    nodes near coords,
  ]
  \addplot coordinates {(100,Frequency)(72,Duration)(21,Motion)};
  \legend{\% of apps}
  \end{axis}
\end{tikzpicture}
\caption{Collection technique distribution}
\label{fig:evidence2}
\end{minipage}\par
\end{figure}

Figs.~\ref{fig:evidence1} and \ref{fig:evidence2} display the distribution of interaction data types and collection techniques identified in our static analysis of the apps' code. The analysis shows that app presentation data and binary data, such as screen content and button clicks, are commonly collected. Moreover, the collection of user input data, particularly in relation to user preferences and surveys, is a frequent practice.

In terms of collection techniques, frequency and duration emerge as common methods. Notably, many apps do not disclose duration-based data collection in their privacy policies. This lack of mention further emphasizes the transparency issues in the way apps communicate their data collection practices.

Our analysis identified five categories of apps that most frequently engage in user interaction data collection: \emph{social, entertainment, shopping, gaming}, and \emph{lifestyle}. The extent of data collection in these categories can be attributed to two key factors. First, the intrinsic characteristics of the category, such as social networking and entertainment, necessitate understanding user behavior for the personalization of services. Second, the complexity of functionality in certain categories, like gaming, often requires learning from user interactions to optimize user experiences. Likewise, lifestyle apps might need to track user actions within the app to function effectively.

Note that almost all apps, across categories, engage in some form of user interaction data collection. However, the level of transparency in detailing such practices in their privacy policies varies widely. The majority of these policies lack completeness, indicating a trend of incomplete disclosure about user interaction data collection practices. This highlights the urgent need for more transparent and detailed communication about these practices in app privacy policies.

\subsection{Case Study: In-depth Analysis of Four Popular Apps}

%Following the general overview, we delve into a more detailed exploration by conducting a case study on four popular apps from the German Google Play Store: WetterOnline (a local weather app), Temu (a Chinese e-commerce platform), Poe (a chatbot app developed by Quora), and Plant App (a plant identification app). Except for Temu, all of these apps provide specific services that seemingly should not involve extensive user interaction data collection. The purpose of this case study is to demonstrate how apps, which may be considered benign, can still have non-transparent policy claims and engage in substantial user interaction data collection. Table~\ref{tab:factcheck} provides an examination of their policy claims alongside our fact-checking results based on the static analysis.
We conducted an in-depth case study on four popular apps from the German Google Play Store to evaluate their privacy policies against actual data collection practices. Our selection included WetterOnline, Temu, Poe, and Plant App. Notably, except for Temu, the expected data collection scope for these services should be minimal. Yet, our static analysis revealed discrepancies in policy transparency, particularly in specifying data collection contexts. Table~\ref{tab:factcheck} provides an examination of their policy claims alongside our fact-checking results.

The analysis showed that while Temu was fairly transparent, the other apps were vague, often using broad terms such as ``interaction with our service'', which lacks detail. WetterOnline and Plant App were particularly limited in disclosing their data collection methods, and Poe's policy was almost silent on its data collection specifics. These findings highlight the critical need for clarity in privacy policies, especially since vague policies can mask practices that might lead to user profiling or re-identification, compromising user trust.

Moreover, the categorization of all user interaction data as non-personal and solely for commercial use is concerning. The extensive behavioral data collected could, when linked with unique identifiers, potentially be used to re-identify users. This underscores the urgent need for policies to more accurately reflect data use, aligning with our aim to ensure user privacy and trust in mobile apps.

\begin{table}[!htb]
    \centering
    \caption{Fact-checking data collection claims wrt.\ evidence for 4 popular apps. {\color{red}Red text} means types of user interaction data missing from the privacy policy/collection claims, while {\color{blue}blue text} means undisclosed techniques of collection.}
    \label{tab:factcheck}
    \footnotesize
    \begin{tabularx}{\textwidth}{ass}
        \toprule
        \textbf{App} & \textbf{Policy Claims} & \textbf{Collection Evidence} \\ \midrule
        \begin{turn}{-90}\textbf{Wetter Online}\end{turn} & The goal of usage measurement is to determine the \textcolor{blue}{intensity} of use, the \textcolor{blue}{number of uses} and users of our application, and their \textcolor{blue}{surfing behavior} statistically. The information about the use (..., \textcolor{red}{the site visited}, date and time of your visit. The event-driven data collection ... is triggered by activities such as \textcolor{red}{installation and start of the app}, ..., and \textcolor{red}{in-app purchases} as well as the \textcolor{red}{receipt}, the \textcolor{red}{swipe} and the \textcolor{red}{opening of push-messages} and the \textcolor{red}{opening and updating of the app by means of a dynamic link}. For each of these events the \textcolor{blue}{number of visits}, the number of users triggering the event and, if available, the value of the events is collected.& Interaction Consistency Rate: data type 3/4; collection technique 1/2. \newline Context Consistency Rate: 1/6. \newline Data types: presentation, categorical, binary, user input. \newline Collection techniques: frequency, duration. \newline Context: viewing content, location, search, notification, sensor-based, customization.\\ 
        \midrule
        \begin{turn}{-90}\textbf{Temu}\end{turn} & Online activity data, such as \textcolor{red}{pages or screens you viewed}, \textcolor{blue}{how long} you spent on a page or screen, the \textcolor{red}{website you visited} before browsing to the Service, \textcolor{blue}{navigation paths} between pages or screens, information about your \textcolor{red}{activity on a page or screen}, \textcolor{blue}{access times and duration} of access, and whether you have \textcolor{red}{opened our emails} or \textcolor{red}{clicked links} within them.& Interaction Consistency Rate: data type 3/5; collection technique 3/3. \newline Context Consistency Rate: 1/10. \newline Data types: presentation, categorical, binary, user input, gesture. \newline Collection techniques: frequency, duration, motion. \newline Context: viewing content, purchase, location, media, search, notification, user profile, sensor-based, communication, customization.\\ \midrule
        \begin{turn}{-90}\textbf{Poe}\end{turn} & Our third party LLM providers and third party bot developers may receive details about your interactions with Poe (including the \textcolor{red}{contents of your chats, upvotes}, etc.) to provide and generally improve their services, which they may process in their legitimate business interests.& Interaction Consistency Rate: data type 1/3; collection technique 0/2. \newline Context Consistency Rate: 1/6. \newline Data types: presentation, binary, user input. \newline Collection techniques: frequency, duration. \newline Context: viewing content, location, search, notification, communication, customization.\\ \midrule
        \begin{turn}{-90}\textbf{PlantApp}\end{turn} & During your visits, we may use software tools such as JavaScript to measure and collect session information including page \textcolor{blue}{response times}, download errors, \textcolor{blue}{length of visits} to certain pages, page interaction information (such as \textcolor{red}{scrolling, clicks}, and \textcolor{blue}{mouse-overs}), and methods used to browse away from the page.& Interaction Consistency Rate: data type 2/5; collection technique 3/3. \newline Context Consistency Rate: 1/8. \newline Data types: presentation, categorical, binary, user input, gesture. \newline Collection techniques: frequency, duration, motion. \newline Context: viewing content, purchase, location, media, search, notification, sensor-based, customization.\\
         \bottomrule
    \end{tabularx}
\vspace{-5mm}
\end{table}

%One striking observation is the lack of specificity in the contexts of data collection described in the policies of these apps. Many use a vague term such as ``when you interact with our service''. Interestingly, although one might suspect Temu, an e-commerce app, to collect a significant amount of user interaction data, our analysis confirms this suspicion but also reveals that Temu's privacy policy is relatively transparent about their data collection practices.

%In contrast, the simpler apps WetterOnline and Plant App provide only limited information regarding their data collection types and techniques. This limitation is even more pronounced in the case of the Poe chatbot, which offers almost negligible information related to user interaction data collection. These examples underscore the importance and need for more transparent claims regarding user interaction data collection in privacy policies.

%Furthermore, we noted that many privacy policies label their user interaction data collection as automatic and accumulative, associating it with an anonymized identifier and stating it is used exclusively for commercial purposes. They often categorize this data as non-personal. However, the vast amount of automatically collected behavioral data, when combined with the collected event-specific values and a unique machine-generated identifier, raises questions about whether such combined data can indeed be classified as non-personal. This issue underlines the need for a more nuanced understanding and categorization of user interaction data in privacy policies.

\section{Related Work}

The analysis of mobile app privacy policies, particularly focusing on user interaction data, forms the core of our research, extending beyond the typical scope of existing studies. While prior research employing NLP techniques has significantly contributed to understanding privacy policies~\cite{10.1145/3180445.3180447,ravichander2021breaking}, our work uniquely concentrates on the nuanced aspects of user interaction data collection. Tools like PrivacyFlash Pro~\cite{zimmeck2021privacyflash} and AutoCog~\cite{qu2014autocog} have laid the groundwork in aligning privacy policy claims with app behaviors, but their focus on personal data leaves a gap in addressing user interaction data, which our study aims to fill.

In the field of static analysis for app security and privacy~\cite{avdiienko2015mining,enck2014taintdroid,zhang2020does}, existing efforts have primarily analyzed app bytecode for data leaks and privacy violations without specifically targeting user interaction data. Our research contributes to this field by introducing an automated method that not only evaluates privacy policy disclosures but also correlates them with actual app behaviors concerning user interaction data. This approach not only enhances transparency but also fosters trust in the mobile app ecosystem, addressing a critical area that has been previously overlooked. Our work, thus, adds a dimension to the current understanding of mobile app privacy and automated data collection practices.

\section{Conclusion and Future Work}

%Through this research, we have examined the practices of user interaction data collection in mobile applications with a focus on transparency. Our automated approach has enabled a direct comparison between privacy policy claims and the actual implemented data collection activities, as identified through static analysis. The findings underscore the need for enhanced transparency and better policy communication in the realm of mobile applications.

%However, the limited scope and the claim-to-evidence mapping based on a self-constructed list of Android UI types present inherent limitations. Future research should expand and refine these methods, ensuring wider coverage of applications, platforms, and analytics services for a comprehensive understanding of data collection practices.

%The common classification of user interaction data as non-personal needs further discussion due to its potential to profile a person when aggregated. Advancing user awareness and fostering responsible practices among developers are essential for a transparent mobile app industry. Our study calls for these actions, contributing to the discourse on responsible and transparent user interaction data collection practices.

Through this research, we investigated the collection of user interaction data in mobile apps, often claimed as anonymized but raising privacy concerns. Recent studies suggest that even anonymized data could be re-identified, challenging the idea of complete privacy protection. We developed a method to compare privacy policy statements with app behaviors, highlighting the need for transparent data collection practices. Our initial results demonstrate effectiveness in identifying the gap between stated practices and actual policy claims regarding interaction data collection, a discrepancy that could erode user trust.

Looking forward, expanding the research to include more apps and platforms will deepen our understanding of data collection practices. Future studies should also examine the classification of user interaction data, typically considered non-personal, and its potential impact on user profiling and privacy. 

\subsubsection*{Acknowledgement}
This paper is an extended version of work published in~\cite{tang2023transparency}. This work is part of the Privacy Matters (PriMa) project. The PriMa project has received funding from European Union’s Horizon 2020 research and innovation program under the Marie Skłodowska-Curie grant agreement No. 860315.

 \bibliographystyle{splncs04}
 \bibliography{reference}

\begin{thebibliography}{10}
\providecommand{\url}[1]{\texttt{#1}}
\providecommand{\urlprefix}{URL }
\providecommand{\doi}[1]{https://doi.org/#1}

\bibitem{avdiienko2015mining}
Avdiienko, V., Kuznetsov, K., Gorla, A., Zeller, A., Arzt, S., Rasthofer, S.,
  Bodden, E.: Mining apps for abnormal usage of sensitive data. In: The 37th
  IEEE international conference on software engineering. vol.~1, pp. 426--436.
  IEEE (2015)

\bibitem{crectu2022interaction}
Cre{\c{t}}u, A.M., Monti, F., Marrone, S., Dong, X., Bronstein, M.,
  de~Montjoye, Y.A.: Interaction data are identifiable even across long periods
  of time. Nature Communications  \textbf{13}(1), ~313 (2022)

\bibitem{devlin2019bert}
Devlin, J., Chang, M.W., Lee, K., Toutanova, K.: Bert: Pre-training of deep
  bidirectional transformers for language understanding (2019)

\bibitem{enck2014taintdroid}
Enck, W., Gilbert, P., Han, S., Tendulkar, V., Chun, B.G., Cox, L.P., Jung, J.,
  McDaniel, P., Sheth, A.N.: Taintdroid: an information-flow tracking system
  for realtime privacy monitoring on smartphones. ACM Transactions on Computer
  Systems (TOCS)  \textbf{32}(2),  1--29 (2014)

\bibitem{grunewald2021tilt}
Gr{\"u}newald, E., Pallas, F.: {TILT: A GDPR-aligned transparency information
  language and toolkit for practical privacy engineering}. In: Proceedings of
  the 2021 ACM Conference on Fairness, Accountability, and Transparency. pp.
  636--646 (2021)

\bibitem{leiva2021my}
Leiva, L.A., Arapakis, I., Iordanou, C.: {My mouse, my rules: Privacy issues of
  behavioral user profiling via mouse tracking}. In: Proceedings of the 2021
  Conference on Human Information Interaction and Retrieval. pp. 51--61 (2021)

\bibitem{bp1}
Marda, V.: {Non-personal data: the case of the Indian Data Protection Bill,
  definitions and assumptions} (2020),
  \url{https://www.adalovelaceinstitute.org/blog/non-personal-data-indian-data-protection-bill/},
  (Accessed on 28/11/2023)

\bibitem{qu2014autocog}
Qu, Z., Rastogi, V., Zhang, X., Chen, Y., Zhu, T., Chen, Z.: Autocog: Measuring
  the description-to-permission fidelity in android applications. In:
  Proceedings of the 2014 ACM SIGSAC Conference on Computer and Communications
  Security. pp. 1354--1365 (2014)

\bibitem{ravichander2021breaking}
Ravichander, A., Black, A.W., Norton, T., Wilson, S., Sadeh, N.: {Breaking Down
  Walls of Text: How Can NLP Benefit Consumer Privacy?} In: Proceedings of the
  59th Annual Meeting of the Association for Computational Linguistics and the
  11th International Joint Conference on Natural Language Processing. vol.~1
  (2021)

\bibitem{SeveralE79:online}
Rzhevkina, A.: {Several EU countries banned Google Analytics - here are some
  alternatives}.
  \url{https://www.contentgrip.com/eu-countries-ban-google-analytics/}
  (September 2022), (Accessed on 03/11/2023)

\bibitem{bp2}
Singh, A., Raghavan, M., Chugh, B., Prasad, S.: {The Contours of Public Policy
  for Non-Personal Data Flows in India} (2019),
  \url{https://www.dvara.com/research/blog/2019/09/24/the-contours-of-public-policy-for-non-personal-data-flows-in-india/},
  (Accessed on 28/11/2023)

\bibitem{tang2023transparency}
Tang, F., Østvold, B.M.: Transparency in app analytics: Analyzing the
  collection of user interaction data. In: 2023 20th Annual International
  Conference on Privacy, Security and Trust (PST). pp. 1--10 (2023).
  \doi{10.1109/PST58708.2023.10320181}

\bibitem{10.1145/3180445.3180447}
Tesfay, W.B., Hofmann, P., Nakamura, T., Kiyomoto, S., Serna, J.:
  {PrivacyGuide: Towards an Implementation of the EU GDPR on Internet Privacy
  Policy Evaluation}. In: Proceedings of the Fourth ACM International Workshop
  on Security and Privacy Analytics. p. 15–21. IWSPA '18 (2018)

\bibitem{zhang2020does}
Zhang, X., Wang, X., Slavin, R., Breaux, T., Niu, J.: How does misconfiguration
  of analytic services compromise mobile privacy? In: Proceedings of the
  ACM/IEEE 42nd International Conference on Software Engineering. pp.
  1572--1583 (2020)

\bibitem{zimmeck2021privacyflash}
Zimmeck, S., Goldstein, R., Baraka, D.: {PrivacyFlash Pro: Automating Privacy
  Policy Generation for Mobile Apps.} In: NDSS (2021)

\bibitem{zimmeck2019maps}
Zimmeck, S., Story, P., Smullen, D., Ravichander, A., Wang, Z., Reidenberg,
  J.R., Russell, N.C., Sadeh, N.: Maps: Scaling privacy compliance analysis to
  a million apps. Proc. Priv. Enhancing Tech.  \textbf{2019}, ~66 (2019)

\end{thebibliography}

\appendix
\section{Implementation of Claim Analysis Using BERT}\label{Sec:claimimple}

We opted for BERT over GPT-3 for its bidirectional architecture, enabling a thorough contextual understanding of privacy policies, essential for our analysis. BERT's capacity to analyze both left and right sentence contexts is particularly effective for interpreting complex privacy policy language~\cite{devlin2019bert}.
In our implementation, BERT was tailored to privacy policy language, involving pre-processing steps like tokenization and normalization, and trained on a specialized dataset to identify binary claims and data collection methods. We also employed bigram analysis to recognize common word pairs, augmenting the model's proficiency in interpreting policy language and thereby enhancing its precision and recall.

The model achieved a precision of 95\% and a recall of 98\% for claim extraction. For data types and collection methods, we observed precision and recall rates of 82\% and 74\%, respectively, and for collection techniques, precision and recall stood at 92\% and 78\%, showcasing the model's robust performance.

\section{Code Analysis and Performance Metrics}\label{Sec:evidencemetrics}

We selected 20 popular apps from the German Google Play Store, meticulously identifying each instance of user interaction data collection to establish a ground truth. Our static analysis method was then evaluated against this dataset.

Our method demonstrated high accuracy (91\%), precision (92\%), and recall (79\%), with an overall F1-score of 85.5\%, indicating effectiveness in accurately identifying and classifying user interaction data collection in mobile apps.

\section{Types of User Interaction Data \& Collection Techn.s}\label{Sec:terms}

We identified six types of user interaction data based on our analysis of Android UI widgets: App Presentation Data, Binary Data, Categorical Data, User Input Data, Gesture Data, and Composite Gestures Data. For a detailed explanation of these types, refer to~\cite{tang2023transparency}.
Similarly, our study categorizes collection techniques as Frequency, Duration, and Motion Details. Each technique's specifics are also elaborated upon in~\cite{tang2023transparency}.

\end{document}